\documentclass[lettersize,journal]{IEEEtran}
\usepackage{amsmath,amssymb,amsfonts}
\usepackage{algorithmic}
\usepackage{algorithm}
\usepackage{array}
\usepackage[caption=false,font=normalsize,labelfont=sf,textfont=sf]{subfig}
\usepackage{textcomp}
\usepackage{stfloats}
\usepackage{url}
\usepackage{verbatim}
\usepackage{graphicx}
\usepackage{cite}
\usepackage{hyperref}       % hyperlinks
\usepackage{booktabs}
\usepackage{multirow}
\def\mathbi#1{\textbf{\em #1}}
\usepackage[capitalize]{cleveref}
\crefname{section}{Sec.}{Secs.}
\Crefname{section}{Section}{Sections}
\Crefname{table}{Table}{Tables}
\crefname{table}{Tab.}{Tabs.}

\begin{document}
\title{JoJoNet: Joint-contrast and Joint-sampling-and-reconstruction Network for Multi-contrast MRI}
\author{Lin Zhao, Xiao Chen, Eric Z. Chen, Yikang Liu, Dinggang Shen*, Terrence Chen, and Shanhui Sun* \thanks{* Co-corresponding authors. (e-mail: Dinggang.Shen@gmail.com and shanhui.sun@uii-ai.com)}
\thanks{L. Zhao is with the Department of Computer Science, University of Georgia, Athens, GA 30602.} \thanks{X. Chen, E. Chen, Y. Liu, T. Chen and S. Sun are with United Imaging Intelligence, Cambridge, MA 02140.} \thanks{D. Shen is with School of Biomedical Engineering, ShanghaiTech University, Shanghai 201210, China, and also with Department of Research and Development, Shanghai United Imaging Intelligence Co., Ltd., Shanghai 200030, China. } \thanks{Contribution from L. Zhao was carried out during his internship at United Imaging Intelligence, Cambridge, MA.}}

\maketitle

\begin{abstract}
Multi-contrast Magnetic Resonance Imaging (MRI) generates multiple medical images with rich and complementary information for routine clinical use; however, it suffers from a long acquisition time. Recent works for accelerating MRI, mainly designed for single contrast, may not be optimal for multi-contrast scenario since the inherent correlations among the multi-contrast images are not exploited. In addition, independent reconstruction of each contrast usually does not translate to optimal performance of downstream tasks. Motivated by these aspects, in this paper we design an end-to-end framework for accelerating multi-contrast MRI which simultaneously optimizes the entire MR imaging workflow including sampling, reconstruction and downstream tasks to achieve the best overall outcomes. The proposed framework consists of a sampling mask generator for each image contrast and a reconstructor exploiting the inter-contrast correlations with a recurrent structure which enables the information sharing in a holistic way. The sampling mask generator and the reconstructor are trained jointly across the multiple image contrasts. The acceleration ratio of each image contrast is also learnable and can be driven by a downstream task performance. We validate our approach on a multi-contrast brain dataset and a multi-contrast knee dataset. Experiments show that (1) our framework consistently outperforms the baselines designed for single contrast on both datasets; (2) our newly designed recurrent reconstruction network effectively improves the reconstruction quality for multi-contrast images; (3) the learnable acceleration ratio improves the downstream task performance significantly. Overall, this work has potentials to open up new avenues for optimizing the entire multi-contrast MR imaging workflow.
\end{abstract}


%%%%%%%%% BODY TEXT
\section{Introduction}
\label{sec:intro}

\begin{figure}[t]
  \centering
   \includegraphics[width=1\linewidth]{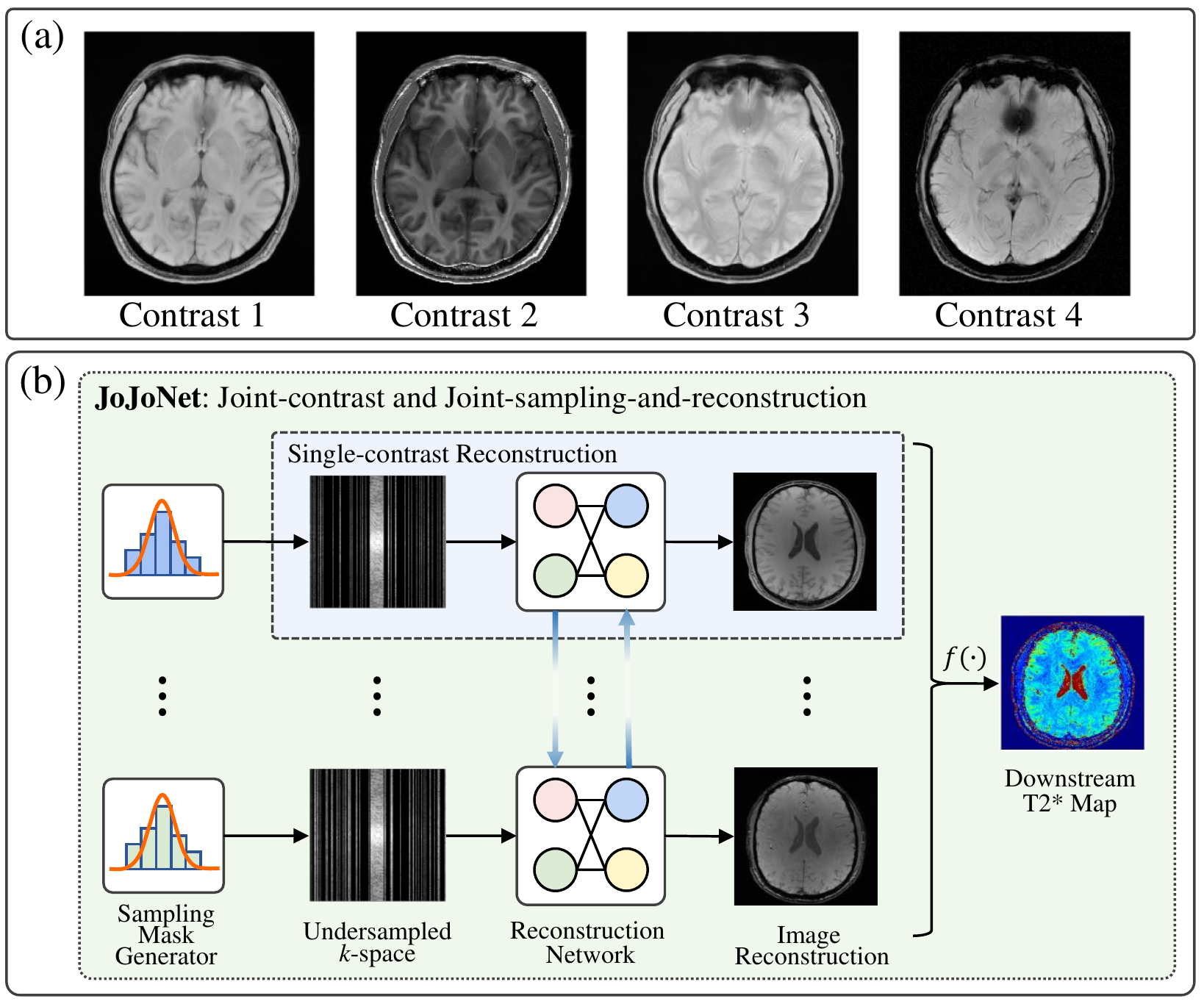}

   \caption{ (a) Example multi-contrast MRI images. (b) Overview of the proposed JoJoNet. Contrast to a single-contrast scheme, the sampling mask generator and reconstruction network are jointly optimized across multiple image contrasts in JoJoNet. The training of the JoJoNet can be driven by individual reconstruction quality or a downstream task performance.}
   \label{fig:figure1}
\end{figure}

\IEEEPARstart{M}{agnetic} Resonance Imaging (MRI) is a widely used and comprehensive medical imaging technique that can offer high-quality images with both anatomical and functional information. For modern MRI nowadays, multi-contrast images with various and distinctive image contrasts (see \Cref{fig:figure1}(a)) are routinely acquired in practice. These multiple contrasts can depict and discriminate different tissues and tissue conditions that reflect the underlying physiological activities, which is commonly used in a complementary manner for diagnosis or as the intermediate images for downstream tasks such as image synthesis \cite{welsch2014t2,taylor2016t1}. On the other hand, MRI is an intrinsically slow imaging modality since data are collected in a point-by-point manner in $k$-space, a complex-valued spatial-frequency space. The slow acquisition not only limits its availability but also causes patient discomfort as well as motion artifacts. The situation becomes even worse for the multi-contrast imaging since it takes longer time to acquire multiple images in a MRI scan. Therefore, accelerating MRI is of decisive significance and has attracted tremendous efforts for decades. 

A common way for accelerating MRI is acquiring only partial $k$-space rather than the full $k$-space \cite{schlemper2017deep,tsao2012mri,zhang2019reducing}. However, undersampling the $k$-space violates the Nyquist-Shannon sampling theorem and results in blurring or aliasing artifacts. Various methods have been proposed to recover high-fidelity and artifact-free images from the undersampled $k$-space, demonstrating great promises especially for those based on deep learning \cite{schlemper2017deep,zhu2018image,bahadir2019learning,zhou2020dudornet,sriram2020grappanet,jun2021joint}. These efforts can be broadly classified into three categories: a) optimizing the sampling pattern that defines the set of $k$-space to be acquired for an easy artifact removal \cite{pineda2020active,bakker2020experimental,sherry2020learning}; b) optimizing the reconstruction method for achieving better image quality \cite{schlemper2017deep,zhou2020dudornet}; c) jointly optimizing the sampling pattern and the reconstruction method \cite{bahadir2019learning,zhang2019reducing,yin2021end}. It is worth noting that the performance of the reconstruction method significantly depends on the sampling pattern \cite{bahadir2019learning, aggarwal2020j}. A joint optimization of the two can adapt the reconstruction method to the specific undersampling pattern and make the undersampling pattern congruent with the reconstruction method, which can further improve reconstruction quality. 

Despite the remarkable progresses and contributions in accelerating MRI, most works focused on accelerating single-contrast imaging while only a few studies considered the multi-contrast scenario \cite{bilgic2011multi, xuan2021multi}. Although images with different contrasts have diversified appearances, correlations among them do exist. These images share the same underlying anatomy and in many MR applications the multiple contrasts record an evolving tissue property where the current contrast depends on the preceding one. In a joint-contrast optimization, these connections among the contrasts can be exploited for sampling and reconstruction design, where data, either acquired or recovered, from one contrast can contribute to another. Furthermore, multi-contrast images are usually used as intermediate images for downstream tasks such as T2* map synthesis where contributions of each contrast are not equal \cite{welsch2014t2}. An optimal strategy would be to collect more data for those critical contrasts and less on the others given a fixed total data amount. Thus, a global optimization of the whole multi-contrast MRI workflow can better distribute the resources among different MR sequences in scanning to achieve the best overall outcomes (image quality and/or downstream task accuracy). 

In this paper, we propose a novel end-to-end framework, joint-contrast joint-sampling-and-reconstruction network (JoJoNet) for accelerating multi-contrast MRI which, for the first time, globally optimizes the entire MR imaging workflow including the $k$-space undersampling, image reconstruction and downstream task performance across multiple contrasts (see \Cref{fig:figure1}(b)). To do so, we learn the individual undersampling pattern for each image contrast and optimize them along with a specially designed reconstruction network, Holistic Recurrent U-Net (HRU-Net), which fully exploits the inter-contrast correlations to produce high-quality multi-contrast reconstructions. Beyond image quality, JoJoNet can also learn to distribute resources (amount of data to be sampled, a.k.a acceleration ratio) across contrasts for optimal downstream task outcomes. The proposed framework is extensively evaluated on a multi-contrast brain dataset and a multi-contrast knee dataset (fastMRI) \cite{zbontar2018fastmri}. The experimental results show that (1) our multi-contrast MRI acceleration framework consistently outperforms those designed for single contrast on both datasets; (2) our HRU-Net demonstrates its effectiveness and superiority in improving the reconstruction quality by utilizing the contrast correlations; (3) in T2* map synthesis downstream task, the scheme with learnable acceleration ratio improves the accuracy of T2* map significantly over the one with fixed acceleration ratio. The contributions of this paper are summarized as follows: 
\begin{itemize}
    \item We introduce a framework for accelerating multi-contrast MRI with global optimization of undersampling and reconstruction across all contrasts.
    \item We propose a novel Holistic Recurrent U-Net to perform multi-contrast reconstruction with the exploitation of inter-contrast correlations.
    \item We optimize MRI downstream task beyond image quality and conduct extensive experiments, demonstrating the proposed method's superiority.
    \item Our study has the potential to open up new avenues for optimizing the entire MR imaging workflow, providing a feasible and effective way for clinical multi-contrast MR applications.
\end{itemize}

\begin{figure*}[t]
  \centering
   \includegraphics[width=1.0\linewidth]{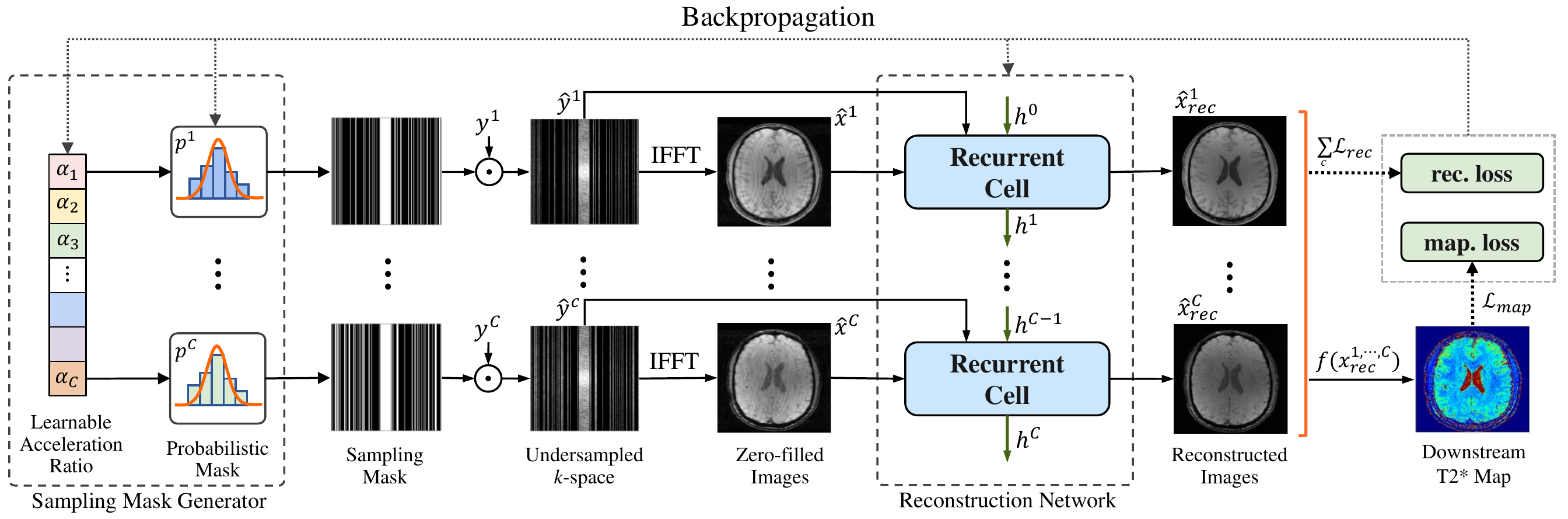}

   \caption{ Illustration of our proposed JoJoNet. The sampling mask generator generates sampling masks to undersample the $k$-space. After IFFT, the resulted \emph{zero-filled} images are fed into reconstruction network which has a recurrent architecture. The reconstructed images can be used for downstream task such as synthesizing T2* map. The training of the whole framework can be driven by either the reconstruction loss or the T2* map loss.}
   \label{fig:figure2}
\end{figure*}

\section{Related works}
\label{sec:literatures}
\textbf{$\mathbi{k}$-space sampling pattern.} A lot of attempts have been made in learning the sampling patterns for undersampling $k$-space. \cite{pineda2020active,bakker2020experimental,jin2019self,zhang2019reducing,xuan2020learning,bahadir2019learning,weiss2020joint,bahadir2020deep,sanchez2020scalable}. For example, reinforcement learning was adopted to learn the policy for actively generating the sampling pattern based on greedy search\cite{bakker2020experimental} and Double Deep Q-Networks \cite{pineda2020active}. \cite{jin2019self} trained a progressive sampler to emulate the policy distribution with a Monte Carlo Tree Search (MCTS). \cite{zhang2019reducing} proposed an evaluator network to perform the active sampling based on
rating the quality gain of each $k$-space measurement in reconstruction. Inspired by the neural network pruning, \cite{xuan2020learning} learned a weight for $k$-space measurements and pruned those with less importance. Other studies turned to the relaxation of binary mask to make the sampling process differentiable \cite{bahadir2019learning,weiss2020joint,bahadir2020deep}. For instance, the recent LOUPE framework \cite{bahadir2019learning,bahadir2020deep} relaxed the binarization of a probabilistic mask with sigmoid function to enable backpropogation. \cite{weiss2020joint} learned a continuous mask to approximate the gradient of its binary version. It is noted that those differentiable approaches can simultaneously optimize the undersampling pattern and reconstruction, improving the reconstruction quality.

\textbf{Image reconstruction.} 
Undersampled MRI reconstruction has been widely studied in the literature. Compressed sensing (CS) based methods incorporated additional a priori knowledge, e.g., sparsity of medical images, to solve the ill-posed reconstruction problem \cite{lustig2007sparse,qu2010iterative,ravishankar2010mr,huang2014bayesian,liu2015balanced, bora2017compressed}. Recently, deep Convolutional Neural Network (CNN) based methods demonstrated their superior performances \cite{wang2016accelerating, schlemper2017deep, zhu2018image, han2019k, sriram2020end}. \cite{schlemper2017deep} presented a cascaded CNN model with residual connections and a data-consistency layer that ensures data fidelity to improve the reconstruction quality, which was widely adopted in the subsequent studies \cite{zhang2019reducing,sriram2020end}. Another group of studies used U-Net architecture \cite{ronneberger2015u} and its variants as the anti-aliasing network \cite{hyun2018deep, bahadir2019learning,han2019k, bakker2020experimental,weiss2020joint,bahadir2020deep,yin2021end}. Several studies integrated the adversarial loss \cite{goodfellow2014generative} to improve the human perceptual reconstruction quality such as the sharpness \cite{yang2017dagan,mardani2017deep,quan2018compressed,seitzer2018adversarial,lei2020wasserstein}. 

There are a few studies dealing with the multi-contrast MRI reconstruction problem\cite{bilgic2011multi,huang2014fast,ehrhardt2016multicontrast,song2019coupled,dar2020prior,xuan2021multi}. For example, \cite{bilgic2011multi} proposed a reconstruction algorithm based on Bayesian CS. \cite{huang2014fast} reconstructed multiple T1/T2-weighted images of the same anatomy based on a joint regularization of total variation (TV) and group wavelet-sparsity. In \cite{song2019coupled}, dictionary learning was adopted to leverage the correlation between different contrasts. Besides these CS-based methods, a recent study proposed a spatial alignment network to register the reference contrast with the target contrast for better quality \cite{xuan2021multi}. However, all the aforementioned approaches only focused on the reconstruction. The joint sampling and reconstruction optimization and the consideration of multi-contrast downstream task performance are still missing.

\section{Background and notation}
\label{sec:bg&not}

Let $y\in\mathbb{C}^{N\times N}$ represents the complex-valued fully-sampled $k$-space. The image $x\in\mathbb{C}^{N\times N}$ can be reconstructed by applying Inverse Fast Fourier Transform (IFFT) $x=\mathcal{F}^{-1}(y)$, and $y=\mathcal{F}(x)$ where $\mathcal{F}$ is the Fast Fourier Transform (FFT). To accelerate MRI, we only acquire a subset of the full $k$-space by undersampling with the Cartesian acquisition trajectory \cite{zhu2018image}. The undersampled $k$-space $\hat{y}$ can be defined as $\hat{y} =y\odot M$, where $M$ is a binary sampling mask determining the sampling pattern, and $\odot$ represents the element-wise product. The \emph{zero-filled} image reconstruction $\hat{x}$ is obtained by $\hat{x}=\mathcal{F}^{-1}(\hat{y})$, which has blurring or aliasing artifacts due to undersampling. An anti-aliasing/denoising function $A(\cdot)$ is applied to get the final reconstructed image $\hat{x}_{rec}=A(\hat{x})$.

We parameterize the anti-aliasing function $A(\cdot)$ as a neural network $A_{\theta}(\cdot)$, i.e., the image reconstruction network. It is noted that different sampling patterns have a huge impact on the types of aliasing artifacts, and determine the performance of reconstruction network implicitly. Thus, we aim to learn a binary sampling mask $M_{p}^{c}$ for each image contrast to collaborate with the reconstruction network for better reconstruction performance. The multi-contrast MRI reconstruction problem is then formulated as the joint optimization of sampling and reconstruction network across multiple image contrasts:
\begin{equation}
  \mathop{\arg\min}_{\theta, p} \ \  \sum_{c=1}^{C}\mathcal{L}_{rec}(A_{\theta}(\mathcal{F}^{-1}(\mathcal{F}(x^{c})\odot M^{c}_{p}),x^{c})
  \label{eq:eq1}
\end{equation}
where $C$ is the number of image contrasts, $x^{c}$ is the $c^{th}$ contrast MR image and $\mathcal{L}_{rec}$ is the loss function measuring reconstruction quality such as the mean squared error (MSE) or structural similarity index measure (SSIM) \cite{wang2004image}.

\begin{figure*}[t]
  \centering
   \includegraphics[width=1.0\linewidth]{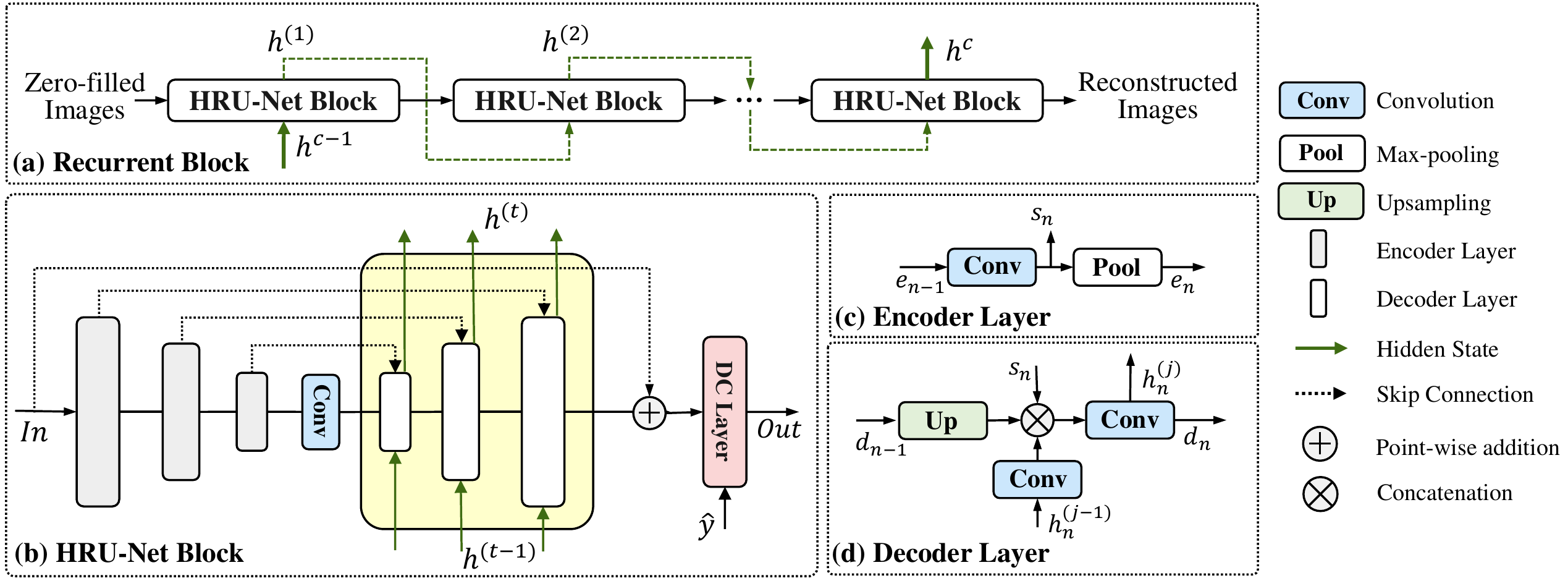}

   \caption{The proposed reconstruction network. (a) Recurrent block which stacks several (b) Holistic Recurrent U-Net (HRU-Net) blocks. The HRU-Net is based on U-Net model. We keep the (c) encoder layer but modify the architecture of decoder (highlighted in yellow box) such that feature maps in all (d) decoder layers are transferred holistically. }
   \label{fig:figure3}
\end{figure*}

\section{Method}
\label{sec:method}
\Cref{fig:figure2} illustrates our approach. The framework consists of a sampling mask generator and a reconstruction network. The sampling mask generator aims to generate the binary sampling masks for undersampling $k$-space for each image contrast, respectively. The reconstruction network takes the \emph{zero-filled} images and the undersampled $k$-space as inputs, and produces the high-fidelity multi-contrast reconstructions. These reconstructions can be the inputs to a downstream task, formulated as a function $f(\cdot)$. In the task of T2* map synthesis, the output of $f(\cdot)$ is the T2* map. The sampling mask generator and the reconstruction network are jointly optimized by image reconstruction quality (rec. loss in \Cref{fig:figure2}) and/or downstream task performance (map. loss in \Cref{fig:figure2}).

\subsection{Sampling mask generator}
For a joint optimization, the sampling mask generator should be differentiable to enable the backpropagation within the whole framework. To do so, we extend the sampling optimization methods in \cite{bahadir2019learning,bahadir2020deep} to be compatible with multi-contrast scenario:
\begin{equation}
\begin{split}
  &\mathop{\arg\min}_{\theta, p} \ \ \frac{1}{C}\sum_{c=1}^{C}\mathcal{L}_{rec}(\\
  &A_{\theta}(\mathcal{F}^{-1}(\mathcal{F}(x^{c})\odot rep(\sigma_s(u^{c} \leq p^{c})))),x^{c}), \\
  &\ \ \ \ \ \ \ \ \ \ \ \ \ \ \ \ \ \ \ \ \ \ \ \ \ \ \ \ \ \ \ \ \ \ \ \ \ \ \ \ \ \ \ \ \ \ \ \ \ \ s.t.\ \  \frac{1}{d}\|p^{c}\|=\alpha
    \label{eq:eq2}
\end{split}
\end{equation}
%\begin{align}
%\mathop{\arg\min}_{\theta, p}\frac{1}{C}\sum_{c=1}^{C}\mathcal{L}_{rec}(A_{\theta}(\mathcal{F}^{-1}(\mathcal{F}(x^{c})\odot rep(\sigma_s(u^{c} \leq p^{c})))),x^{c}), & \nonumber\\ 
%s.t. \frac{1}{d}\|p^{c}\|=\alpha &
%    \label{eq:eq2}
%\end{align}
$u^{c}\in\mathbb{C}^{N}$ is a realization of a random vector that is uniformly distributed on $[0,1]$. $p^{c}\in\mathbb{C}^{N}$ is a ``probabilistic mask" and is binarized by $u^{c} \leq p^{c}$ which sets the value to 1 if the inequality is satisfied and 0 otherwise. $rep(u^{c} \leq p^{c})$ expands the dimension and replicates the elements to make a 2D binary matrix $rep(u^{c} \leq p^{c}) \in\mathbb{C}^{N\times N}$. However, this binarization process is not differentiable. To enable backpropagation, the binarization is relaxed with sigmoid function $\sigma_s(u^{c} \leq p^{c})$, where $\sigma_s(t)=1/(1+e^{-st})$ and $s$ is the slope. In the forward pass, the original binarization function is used but in the backpropagation, the gradient of the relaxed function is used for approximation \cite{bahadir2019learning,bahadir2020deep}. To make the binary mask satisfy the desired sparsity $\alpha$ (acceleration ratio $r=1/\alpha$), a constraint $\|p^{c}\|/ d=\alpha$ is added, where $d$ is the total number of columns in $k$-space. In this way, we enable the joint-contrast and joint-sampling-and-reconstruction optimization for multi-contrast images.
%-------------------------------------------------------------------------
\subsection{Learnable acceleration ratio}
\label{sec:learnable}
For the multi-contrast MRI, the importance of each image contrast could be different, especially in some downstream tasks such as T2 and T2* map synthesis \cite{welsch2014t2}. We aim to learn the acceleration ratio for different contrast automatically such that the resources (scanning time) can be better distributed in terms of downstream task performance. Specifically, we define a learnable parameter $\boldsymbol{w}\in\mathbb{R}^{C}$. The sparsity for each contrast image $\alpha_c$ is calculated via the \emph{Sofrmax} function:

\begin{equation}
\alpha_c = \sigma(\boldsymbol{w})_c=\alpha\times\frac{e^{w_c}}{\sum_{i=1}^C {e^{w_i}}}
  \label{eq:learnable_acc}
\end{equation}
where $\alpha=1/r$ is the overall sparsity and $r$ is the overall acceleration ratio. The constraint in \Cref{eq:eq2} is then modified as $\|p^{c}\|/d=\alpha_c$. In this way, the acceleration ratio for each contrast can be optimized together with the sampling mask generator and the reconstruction network.

\subsection{Reconstruction network}
As illustrated in \Cref{fig:figure2}, our reconstruction network has a recurrent architecture. During training, for the $c^{th}$ image contrast, the binary mask generated by the sampling mask generator is firstly applied to the fully-sampled $k$-space $y^c$ to obtain the undersampled $k$-space $\hat{y}^c$, mimicking MR acquisition process. After IFFT, the resulting \emph{zero-filled} reconstructed image is input into our carefully designed Recurrent Cell along with the hidden state $h^{c-1}$ from previous contrast. Then the Recurrent Cell outputs the high-fidelity reconstructed image for the current contrast and the hidden state $h^{c}$ for the next contrast. The rationale behind utilizing the recurrent architecture is that different image contrasts have correlations such as the same or similar anatomy, and the abstraction of these correlations from previous contrasts can benefit the reconstruction for the current contrast.

\Cref{fig:figure3} illustrates our design of the Recurrent Cell. It has a cascaded backbone composed of several Holistic Recurrent U-Net (HRU-Net) blocks which includes a HRU-Net and a data consistency (DC) layer \cite{schlemper2017deep}. The HRU-Net is newly designed based on U-Net \cite{ronneberger2015u} but several key modifications in the decoder make it compatible with the recurrent input. Specifically, in the $n^{th}$ decoder layer of the $j^{th}$ HRU-Net block, the upsampling of the feature maps $d_{n-1}$ from preceding decoder layer, the feature maps $s_n$ from corresponding encoder layer via skip connection and the convolution of the input hidden state $h^{(j-1)}_{n}$ are concatenated together for convolution operation. The output after convolution serves as the hidden state $h^{(j)}_{n}$ for the next block at decoder layer $n$ as well as the input $d_{n}$ for the following decoder layer. The rationale behind this design is that feature maps from all decoder layers of the current block will be directly utilized in the decoding of the next block , which maximally retains the shareable information and reduces the information loss in both encoding and decoding process of the U-Net. Compared with other recurrent U-Net models such as \cite{wang2019recurrent} which only passes the information in the bottleneck for image segmentation, our design transmits the information in all decoder layers (highlighted in yellow box in \Cref{fig:figure3}). Thus, we named it “Holistic Recurrent U-Net”. In addtion, the direct skip connection from the input of HRU-Net to its output is built to enforce the HRU-Net on learning the residuals. 

The DC layer in the network ensures the estimated image is consistent with the acquired data \cite{schlemper2017deep}. Specifically, in the DC layer FFT is firstly applied to the reconstructed image $\hat{x}_{rec}$ from the HRU-Net to obtain the predicted $k$-space. Then the undersampled $k$-space $\hat{y}$ replaces data at the same sampling locations in the predicted $k$-space. The DC operation can be formally defined as:
\begin{equation}
  DC(M,\hat{x}_{rec},\hat{y}) = \mathcal{F}^{-1}((1-M)\odot\mathcal{F}(\hat{x}_{rec})+M\odot\hat{y} )
  \label{eq:recon_target}
\end{equation}

By stacking the HRU-Net block, the cascade can have $N_l$ blocks, and we investigate the effects of $N_l$ in reconstruction performance in \Cref{sec:nleffects}.

\subsection{Downstream task optimization}
We also consider the downstream task performance in the joint optimization. In this study we use the T2* map synthesis as an example. On the T2*-weighted multi-contrast images, the pixel value on the $c^{th}$ image at location $(u,v)$ is $s(u,v,c)=e^{-c\cdot\Delta t/T2^{*}(u,v)}$. Assuming $f(\cdot)$ is a function that maps $\{s(u,v,1), s(u,v,2), ..., s(u,v,C)\}$ to $T2^{*}(u,v)$, a complete T2* map is synthesized after $f(\cdot)$ is applied at all pixel locations. We train a separate neural network to perform $f(\cdot)$ and then fix the parameters. Ground truth T2* maps are generated from fully-sampled images. A T2* map synthesis optimization is then formulated as below in the learnable acceleration ratio scenario:
%\begin{equation}
%\begin{split}
%  &\mathop{\arg\min}_{\theta,p} \ \ \mathcal{L}_{map}(\\
%  &f(\{A_{\theta}(\mathcal{F}^{-1}(\mathcal{F}(x^{c})\odot M^{c}_p))\}),f(\{x^{c)
%  }\}), \\
%  &\ \ \ \ \ \ \ \ \ \ \ \ \ \ \ \ \ \ \ \ \ \ \ \ \ \ \ \ \ \ \ \ \ \ \ \ \ \ \ \ \ \ \ \ \ \ \ \ \ \ s.t.\ \  \frac{1}{d}\|p^{c}\|=\alpha_c
%    \label{eq:eq4}
%\end{split}
%\end{equation}
%\begin{gather}
\begin{align}
  \mathop{\arg\min}_{\theta,p}  \mathcal{L}_{map}(
  f(\{A_{\theta}(\mathcal{F}^{-1}(\mathcal{F}(x^{c})\odot M^{c}_p))\}),f(\{x^{c)
  }\}) & \nonumber \\
   s.t.  \frac{1}{d}\|p^{c}\|=\alpha_c, &
    \label{eq:eq4}
\end{align}
%\end{gather}
where $\{x^{c}\}=\{x^{1},x^{2},\cdots,x^{C}\}$ and $\mathcal{L}_{map}$ is the loss function on maps such as MSE.

\section{Experiments}
\label{sec:experiment}

\textbf{Dataset.} Our experiments are conducted on a private multi-contrast 3D brain MRI dataset and a public multi-contrast 3D knee dataset (fastMRI) \cite{zbontar2018fastmri}. The brain dataset has 66 subjects, each of which has five different T2*-weighted contrasts. We firstly divide the dataset at patient level and then slice the volumes in the transverse section, resulting in 3644 training slices, 584 validation slices and 584 slices for testing. We report the results on the test dataset. All slices are resized with a resolution 256$\times$224 and are normalized by the image mean.

The fastMRI knee dataset has two contrasts, proton-density (PD) and proton-density with fast suppression (PDFS). More details are referred to \cite{zbontar2018fastmri}. We extract slices along the coronal direction, obtaining 14970/3289/3289 slices for training/testing/validation. Our experimental results are based on the testing. We resize each image to resolution 320$\times$320 and normalize it by the image mean.
Complex-valued multi-coil data are used for all experiments. The multi-coil data are handled in the DC layer using coil sensitivity maps as in ~\cite{hammernik2018learning}.

\textbf{Evaluation metrics.} In our experiments, the reconstruction quality is evaluated by peak signal to noise ratio (PSNR) and structural similarity index measure (SSIM) \cite{wang2004image} between the reconstructed images/maps and ground truth images/maps. For the maps, metrics calculated using and without using a background mask are both reported. 

\textbf{Implementation details.} In all experiments, we pre-select 20 columns of $k$-space measurements in the central low spatial-frequency band and make the sampling mask generator generate the rest. All U-Net based models have 4 encoding/decoding layers with 64/128/256/512 filters in each layer, respectively. The recurrent cell has 3 HRU-Net blocks. The framework is implemented with PyTorch \cite{paszke2019pytorch} deep learning library. We used the Adam optimizer \cite{kingma2014adam} with $\beta_1=0.9$ and $\beta_2=0.999$. The batch size for multi-contrast brain dataset and knee dataset is 16/4, respectively. The model is trained for 100 epochs with a learning rate 0.01 on two NVIDIA A100 GPUs.

\subsection{Comparisons of reconstructions}

In this subsection, we compare the proposed JoJoNet reconstruction framework with several baselines: (a) RM-S/U-Net-S: a shared random sampling mask as in \cite{zbontar2018fastmri} and a shared U-Net reconstruction network \cite{ronneberger2015u} for all image contrasts; (b) LM-S/U-Net-S: a shared learnable sampling mask as in LOUPE \cite{bahadir2019learning} and a shared U-Net reconstruction network for all contrasts; (c) LM-I/U-Net-I: individual learnable sampling mask and individual U-Net reconstruction network for each contrast; (d) LM-I/U-Net-S: individual learnable sampling mask for each contrast and a shared U-Net reconstruction network for all contrasts; (e) LM-I/GRU-U-Net: individual learnable sampling mask for each contrast and a reconstruction network based on Gated Recurrent Unit (GRU) model \cite{cho2014learning} using the reconstructed image from previous contrast as hidden state. The convolution of the hidden state and the \emph{zero-filled} image are concatenated and input into a U-Net for image reconstruction; (f) LM-I/RU-Net: individual learnable sampling mask for each contrast and a reconstruction network based on Recurrent U-Net model \cite{wang2019recurrent} that only passes the information in the bottleneck. Note that baselines d-f all fall into the JoJoNet framework and (g) LM-I/HRU-Net is the one with the proposed HRU-Net for reconstruction.

\begin{table}[h]
  \centering
  \caption{The averaged PSNR (dB) and SSIM for all compared baselines and the proposed LM-I/HRU-Net model over all contrasts on two different MRI datasets. Abbreviation: RM: Random Mask; LM: Learnable Mask; -S: Shared; -I: Individual.}
  \begin{tabular}{lcccc}
    \toprule
    \multicolumn{1}{c}{\multirow{2}{*}{Methods}} &  \multicolumn{2}{c}{Brain} & \multicolumn{2}{c}{Knee}\\
     & PSNR & SSIM & PSNR & SSIM\\ 
    \midrule
    (a) RM-S/U-Net-S & 30.94 & 0.8226 & 33.88 & 0.8570 \\
    (b) LM-S/U-Net-S & 31.54 & 0.8434 & 35.14 & 0.8772  \\
    (c) LM-I/U-Net-I & 31.44 & 0.8412 & 35.20 & 0.8775 \\
    (d) LM-I/U-Net-S & 31.53 & 0.8424 & 35.22 & 0.8777 \\
    (e) LM-I/GRU-U-Net & 31.84 & 0.8481 & 35.29 & 0.8789 \\
    (f) LM-I/RU-Net & 32.85 & 0.8503  & 36.06 & 0.8853 \\
    (g) \textbf{LM-I/HRU-Net} &  \textbf{33.93} & \textbf{0.8652} & \textbf{36.32} & \textbf{0.8885} \\
    \bottomrule
  \end{tabular}
  \label{tab:table1}
\end{table}

\begin{figure}[h]
  \centering
   \includegraphics[width=1\linewidth]{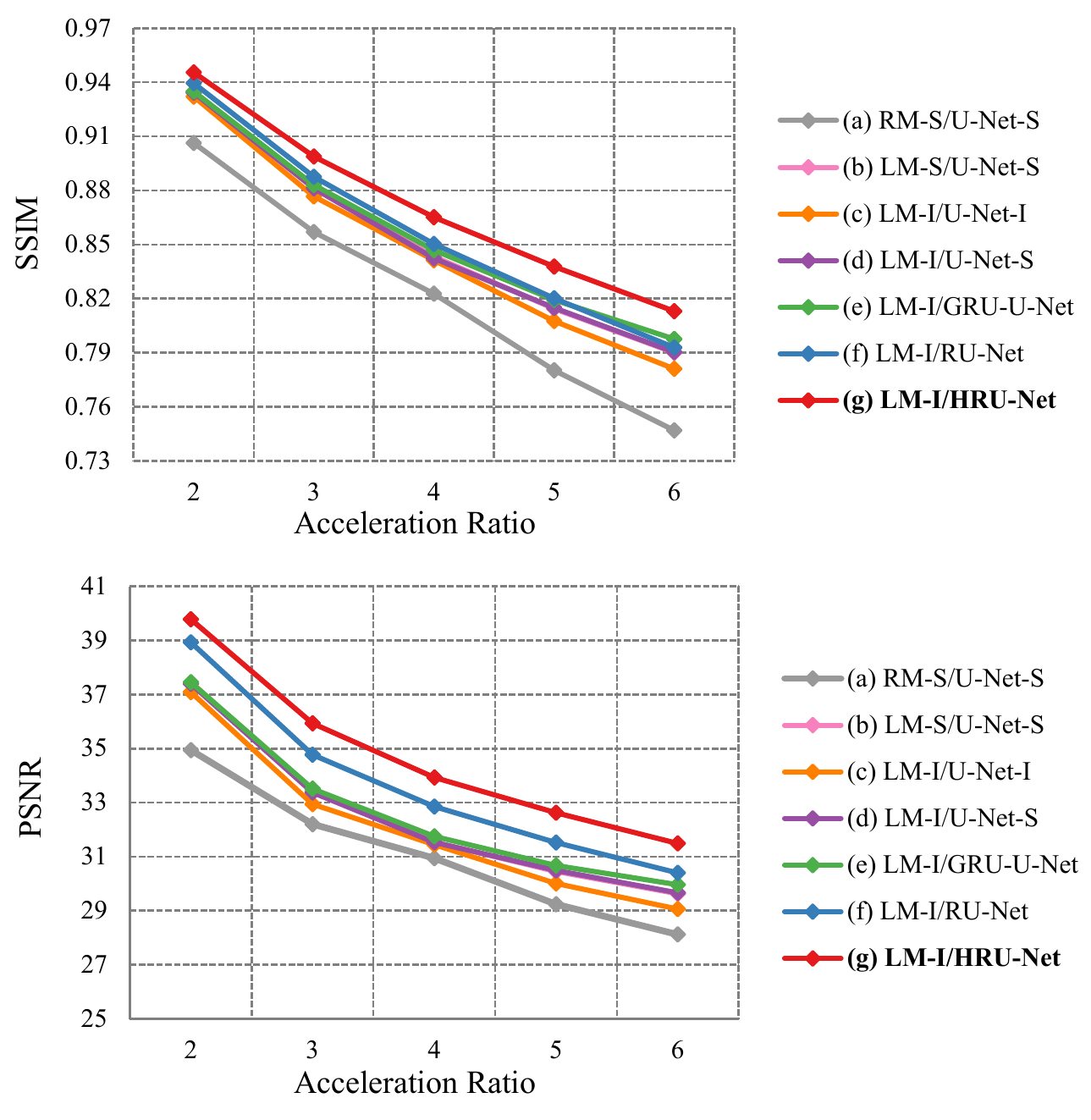}

   \caption{The comparison of averaged PSNR (dB) and SSIM under different acceleration ratios on multi-contrast brain dataset for all compared baslines and the proposed LM-I/HRU-Net.}
   \label{fig:figure4}
\end{figure}

\begin{figure*}[!h]
  \centering
  %\fbox{\rule{0pt}{2in} \rule{1\linewidth}{0pt}}
   \includegraphics[width=1\linewidth]{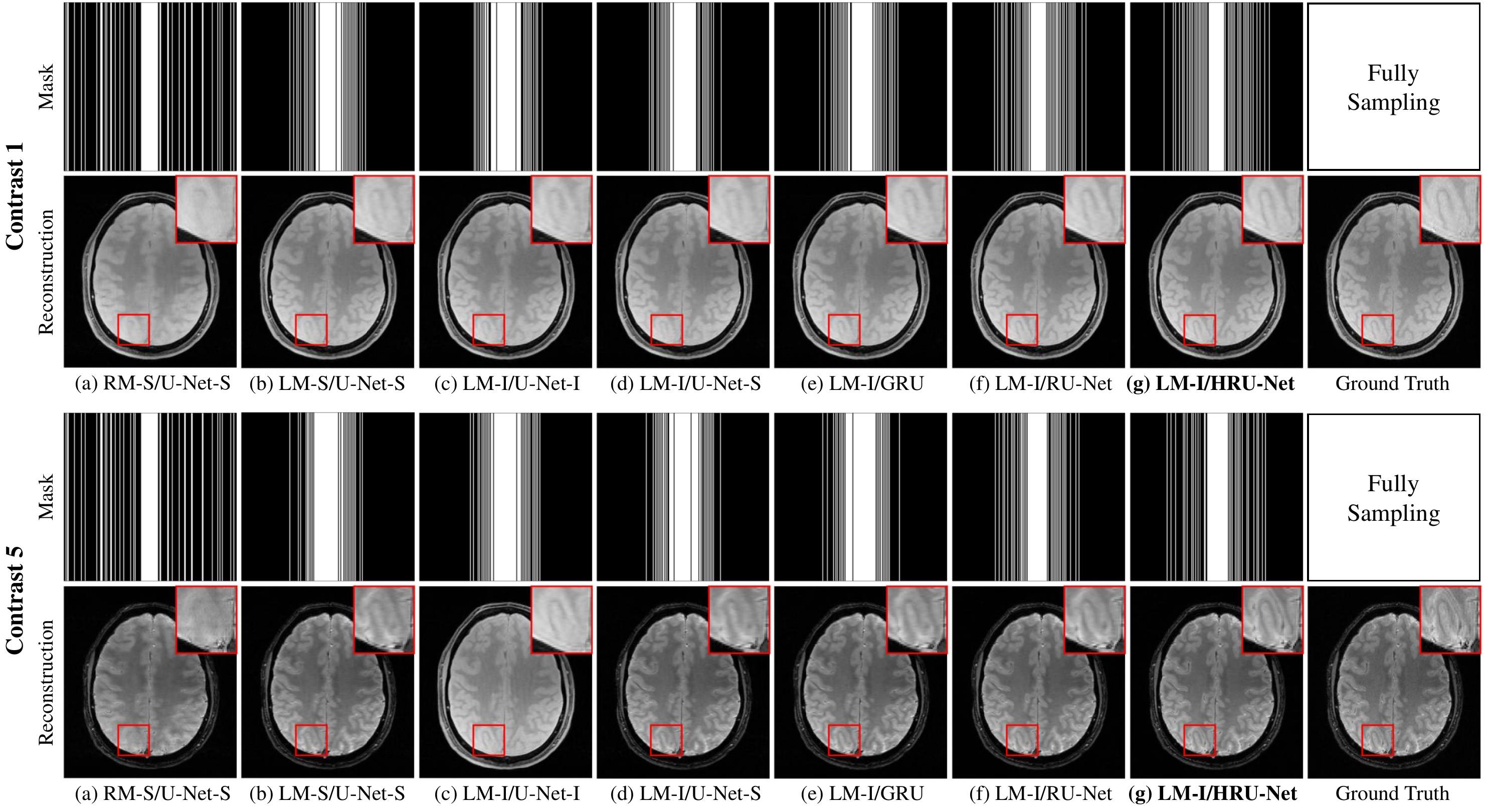}

   \caption{Qualitative comparison of all baselines and the proposed method on multi-contrast brain dataset (4x acceleration), including reconstructed images and sampling masks from two image contrasts.}
   \label{fig:figure5}
\end{figure*}

\begin{figure*}[!h]
  \centering
  %\fbox{\rule{0pt}{2in} \rule{1\linewidth}{0pt}}
   \includegraphics[width=1\linewidth]{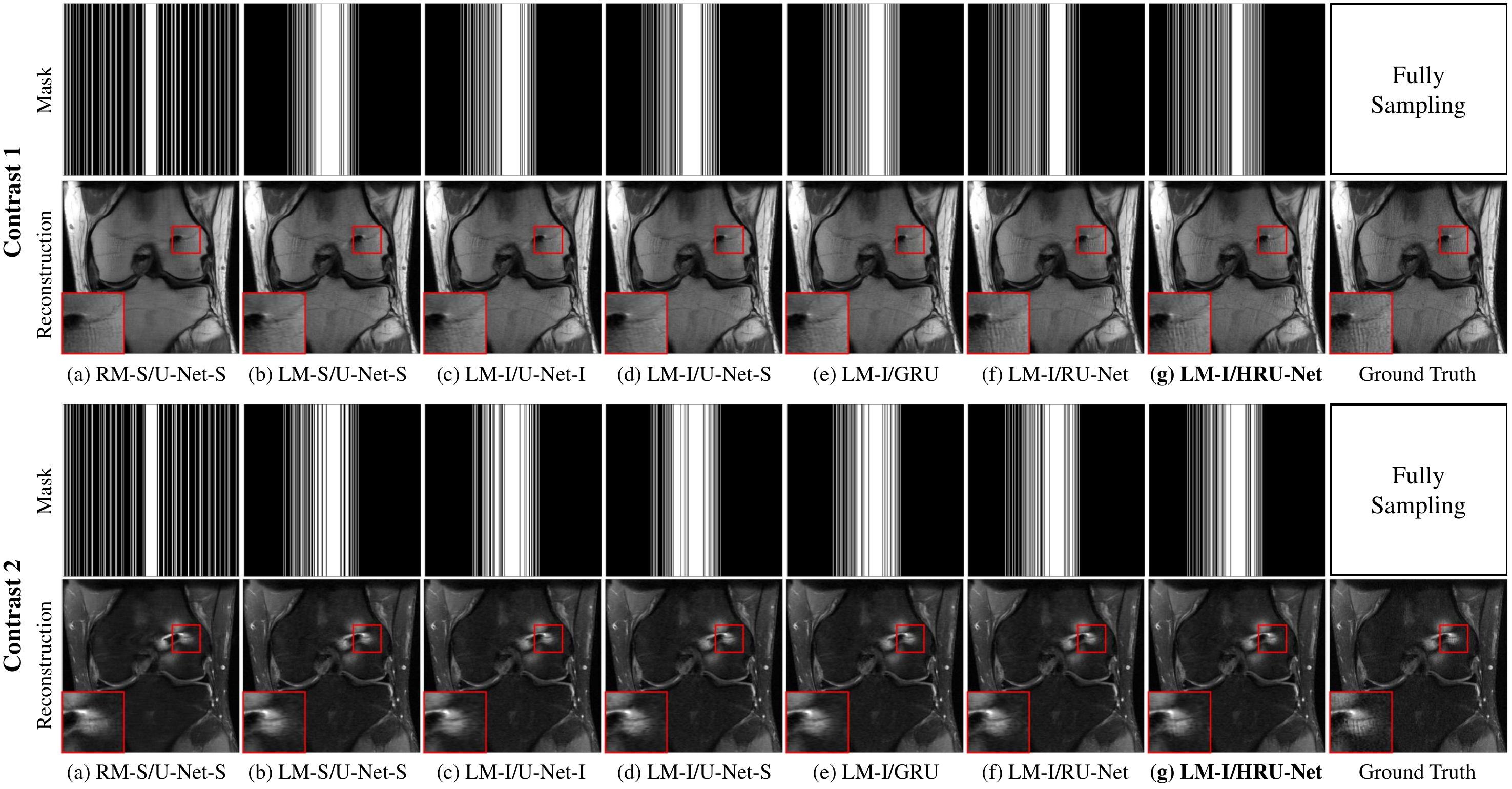}

   \caption{Qualitative comparison of all baselines and the proposed method on multi-contrast knee dataset (4x acceleration), including reconstructed images and sampling masks from two image contrasts.}
   \label{fig:figure6}
\end{figure*}

\Cref{tab:table1} reports the mean PSNR and SSIM over all contrasts for all baselines and our method on the brain and the knee datasets (4x acceleration). The proposed framework achieves state-of-the-art in terms of PSNR and SSIM on both datasets. Our method outperforms all baselines for all image contrasts.

Compared with (a) RM-S/U-Net-S, i.e., the one with random mask, (b) LM-S/U-Net-S with learnable sampling improves the reconstructed image significantly, demonstrating the superiority of jointly optimizing the sampling and the reconstruction, which also agrees with previous single-contrast studies ~\cite{bahadir2019learning}. Baseline (b) LM-S/U-Net-S and (d) LM-I/U-Net-S are comparable. The only difference between them is the baseline (b) learns a shared sampling mask for all contrasts while the baseline (d) learns individual sampling masks for each contrast. The comparable performance is probably due to the limitation of the shared U-Net so the sampling mask generator has to generate the similar sampling pattern for each contrast such that the reconstructor can deal with all contrasts altogether.

We observed that recurrent models ((e)GRU-U-Net, (f)RU-Net, (g)HRU-Net) outperform all other baselines without information transferring among different contrasts, which aligns with our expectation that exploiting the inter-contrast correlation contributes to multi-contrast image reconstruction quality. Baseline (f) LM-I/RU-Net and our method LM-I/HRU-Net have a better performance than (e) LM-I/GRU-U-Net. It may be because the hidden states in GRU model have to undergo the whole encoding and decoding process in U-Net to benefit the reconstruction, which causes the information loss. It is noted that our LM-I/HRU-Net model also exceeds the (f) LM-I/RU-Net with an remarkable improvement under the same settings, e.g., same number of U-Net layers and stacked recurrent block. It shows that our holistic design that transfers the features in all decoding layers is more suitable for reconstruction problem than transferring the features around bottleneck as in RU-Net.

\begin{figure*}[h]
  \centering
   \includegraphics[width=0.95\linewidth]{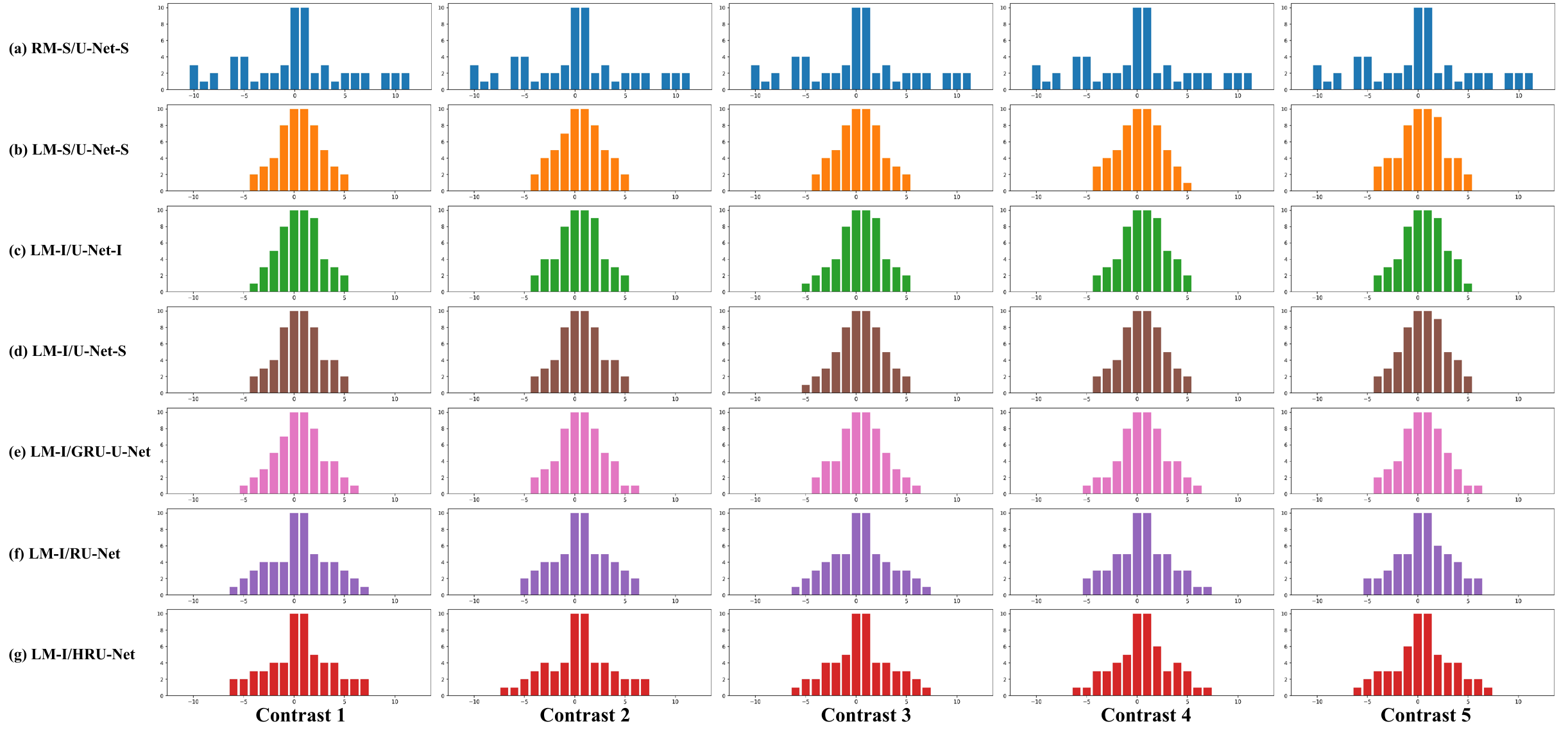}

   \caption{Comparisons of learned mask histogram among different methods on multi-contrast brain dataset (4x acceleration). The $k$-space is divided into small bins with the length of 10 in phase-encoding direction and each bar corresponds to the number of sampled components in each bin.}
   \label{fig:figurem}
\end{figure*}

We also evaluate the performances of all compared baselines under different acceleration ratios (2x to 6x) in \Cref{fig:figure4}. The proposed LM-I/HRU-Net consistently outperforms all compared baselines under all acceleration ratios, indicating the effectiveness and superiority of our framework.

\begin{figure}[ht]
  \centering
   \includegraphics[width=1.0\linewidth]{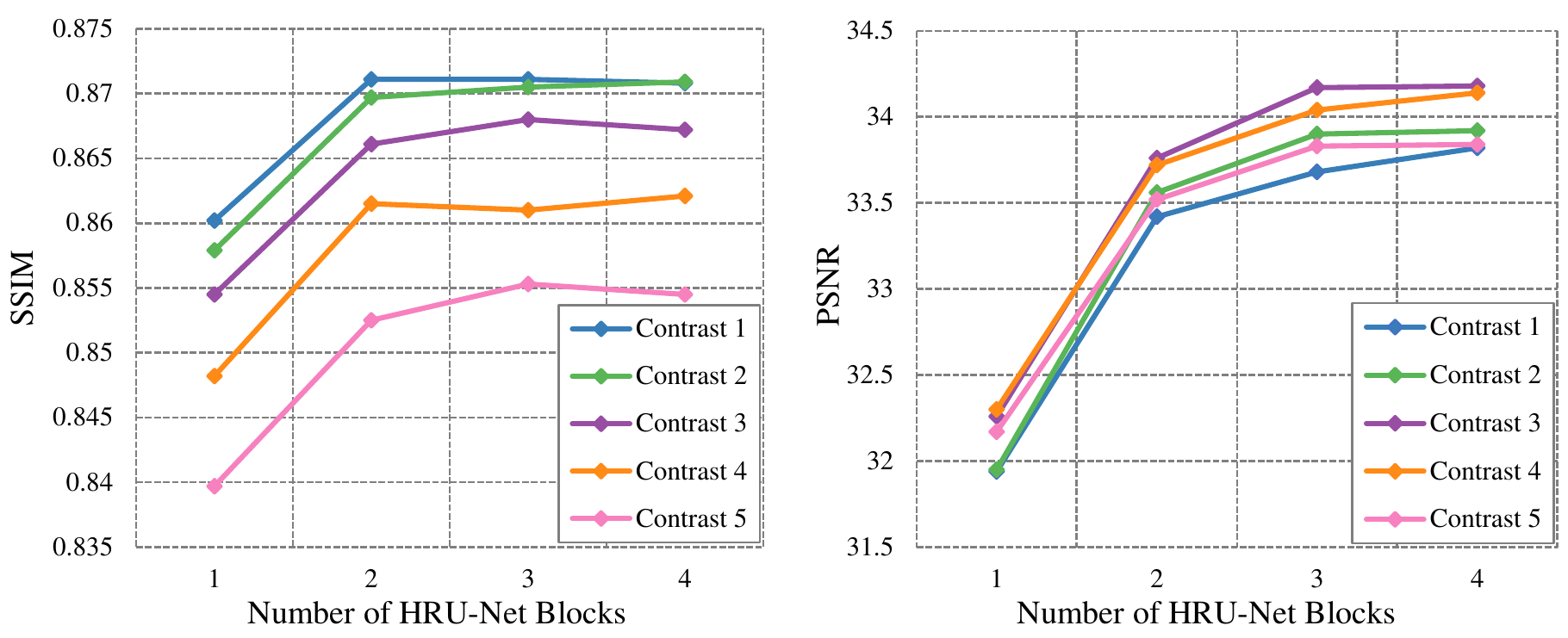}

   \caption{PSNR(dB) and SSIM of the proposed model (4x acceleration) with different numbers of HRU-Net blocks on multi-contrast brain dataset.}
   \label{fig:figure7}
\end{figure}

\begin{table*}[h]
  \centering
  \caption{The total and individual acceleration ratios in T2* mapping task and the PSNR/SSIM for synthesized T2* map with and without background. Results from fixed acceleration ratio and learnable acceleration ratio are shown and compared. Abbreviation: Total Acc: total acceleration ratio; BG: with background; NBG: without background.}  
  \begin{tabular}{cccccccrc}
    \toprule
    Total& \multirow{2}{*}{Methods} & \multicolumn{5}{c}{Individual Acceleration Ratio} & \multicolumn{2}{c}{PSNR(dB) / SSIM} \\
    Acc&& Contrast 1 & Contrast 2 & Contrast 3 & Contrast 4 & Contrast 5 & T2* Map BG & T2* Map NBG\\
    \midrule
    \multirow{2}{*}{2} & Fixed & 2.0 & 2.0 & 2.0 & 2.0 & 2.0 & 9.99 / 0.6311 & 18.67 / 0.8939\\
     & \textbf{Learnable}& 1.0 & 3.3 & 7.3 & 8.4 & 1.0 & \textbf{13.81 / 0.8123} & \textbf{20.12 / 0.9263}\\
    \midrule
    \multirow{2}{*}{4} & Fixed  & 4.0 & 4.0 & 4.0 & 4.0 & 4.0 & 8.40 / 0.4133 & 15.40 / 0.7608\\
     & \textbf{Learnable} & 1.6 & 11.2 & 11.2 & 11.2 & 1.6 & \textbf{11.52 / 0.6157} & \textbf{17.94 / 0.8663}\\
    \bottomrule
  \end{tabular}
  \label{tab:table2}
\end{table*}

We visualize example reconstructed images and the sampling masks in \Cref{fig:figure5} (brain) and \Cref{fig:figure6} (knee) for qualitative comparisons. It is observed that reconstructions from the proposed model have more precise and distinct boundaries between white matter and gray matter (\Cref{fig:figure5}) while those from baselines are over-smoothed and blurry. In \Cref{fig:figure6}, our model recovers more details compared with other baselines.

\subsection{Comparisons of learned masks}
\label{sec:learnedmask}
We divide the $k$-space into bins with a length of 10 in phase-encoding direction and count the number of sampled components in each bin. The averaged number of each bin under 4x acceleration is demonstrated as bar graphs in \Cref{fig:figurem}. The sampling mask learned by our model shows more sampling in the high-frequency bands (edges of the $k$-space). On the other hand, the compared methods with inferior performance have a denser sampling on low-frequency band (center of the $k$-space). This observation is well reproduced with other acceleration ratios. And it is also consistent with results of qualitative comparison in \Cref{fig:figure5} and \Cref{fig:figure6}. Considering that the learnable sampling mask is jointly optimized with reconstruction network, this indicates that in our model more low-frequency information are inherited from previous contrasts thus the sampling mask generator can afford to acquire more high-frequency components to improve the reconstruction quality.

\subsection{Effect of stacking HRU-Net blocks}
\label{sec:nleffects}
In this subsection, we investigated the effect of increasing the number of HRU-Net blocks, $N_{l}$, on the brain dataset. As shown in \Cref{fig:figure7}, the reconstruction performance increases with more HRU-Net blocks. The model with two HRU-Net blocks improves the image quality significantly compared with the one having only one block, while the improvements become saturated when $N_l=3$. In our study, we stack 3 HRU-Net blocks to keep a balance between the reconstruction quality and computing time.

\begin{figure}[h]
  \centering
   \includegraphics[width=0.95\linewidth]{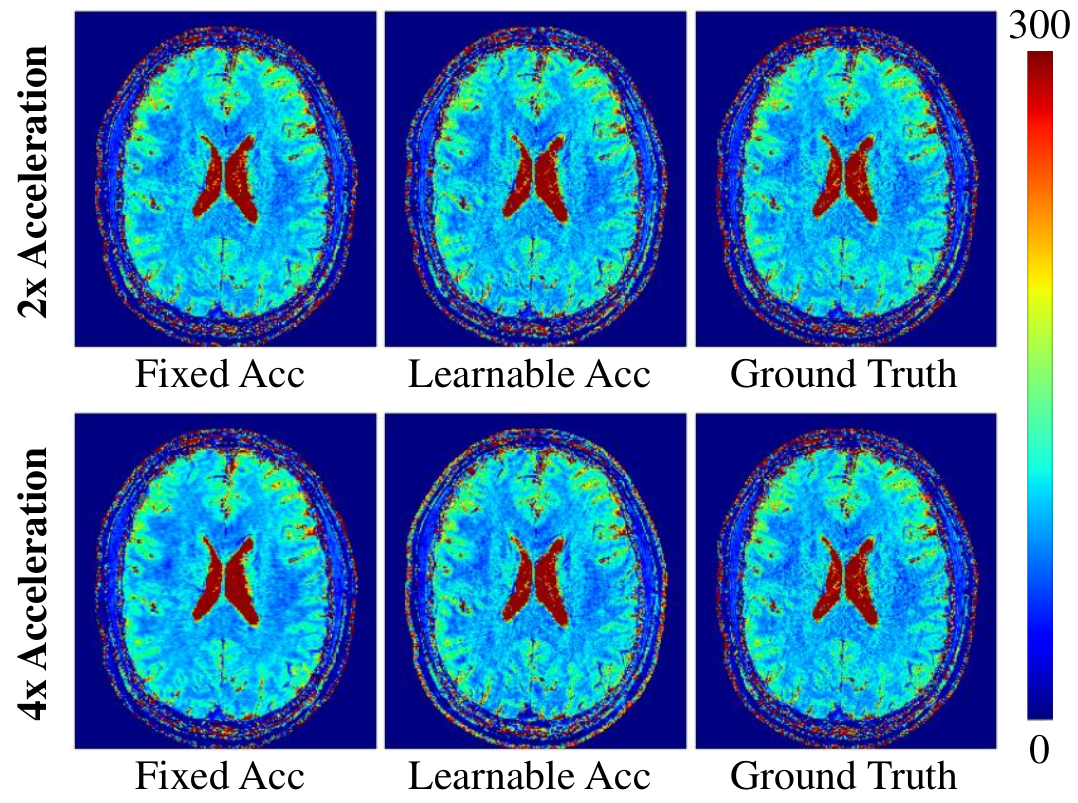}

   \caption{The T2* maps based on the fixed and learnable acceleration ratio under 2x and 4x acceleration settings.}
   \label{fig:figure8}
\end{figure}

\subsection{Learnable acceleration ratio in T2* mapping}
In this subsection, we evaluated the proposed learnable acceleration ratio in T2* map synthesis downstream task on private brain dataset. \Cref{tab:table2} compares T2* map accuracy measured by PSNR and SSIM between the learnable acceleration ratio and the fixed acceleration ratio. The learnable acceleration ratio improves the map quality effectively with the PSNR and the SSIM metrics. It is also observed that the contrast 1 and contrast 5 are granted more k-space measurements, and thus they have better reconstruction quality, in both 2x and 4x accelerations compared with other contrasts. This observation is consistent with the fact that the first and the last contrasts are more important in fitting the exponential decay in T2* mapping task, which provide the largest signal dynamic range for more robustness against noise. For the rest contrasts, early contrasts are favored by the learning. This is also consistent with the fact that early time points in exponential decay captures more signal variations than later ones. We visualize the T2* map from one randomly selected slice in \Cref{fig:figure8}. The T2* map from learnable acceleration provides more accurate T2* values and has sharper anatomy boundaries compared to the one with fixed acceleration. Overall, the experimental results demonstrate the effectiveness of our learnable acceleration ratio in optimizing the multi-contrast downstream tasks.

\section{Conclusions}
In this paper, we presented a novel framework JoJoNet for optimizing the entire multi-contrast MRI workflow, which can jointly generate the sampling pattern, reconstruct high-quality MR images across multiple image contrasts and optimize the downstream task performance. JoJoNet incorporates MR physics deeply into the sampling and the reconstruction optimizations, exploring the correlations among different contrasts. Extensive evaluations show that our JoJoNet outperforms those designed for single contrast MRI. The HRU-Net block demonstrates its superiority in improving the reconstruction quality. The learnable acceleration ratio improves the downstream task performance significantly. Future work includes extending network structure for any order of contrasts and evaluating on other tasks.

%%%%%%%%% REFERENCES
{\small
\bibliographystyle{IEEEtran}
\bibliography{egbib}
}

\end{document}

\title{Supplemental Materials}

\maketitle

\section{Configurations of baseline methods}
\label{sec:baseline}

\begin{figure*}[ht]
  \centering
   \includegraphics[width=1\linewidth]{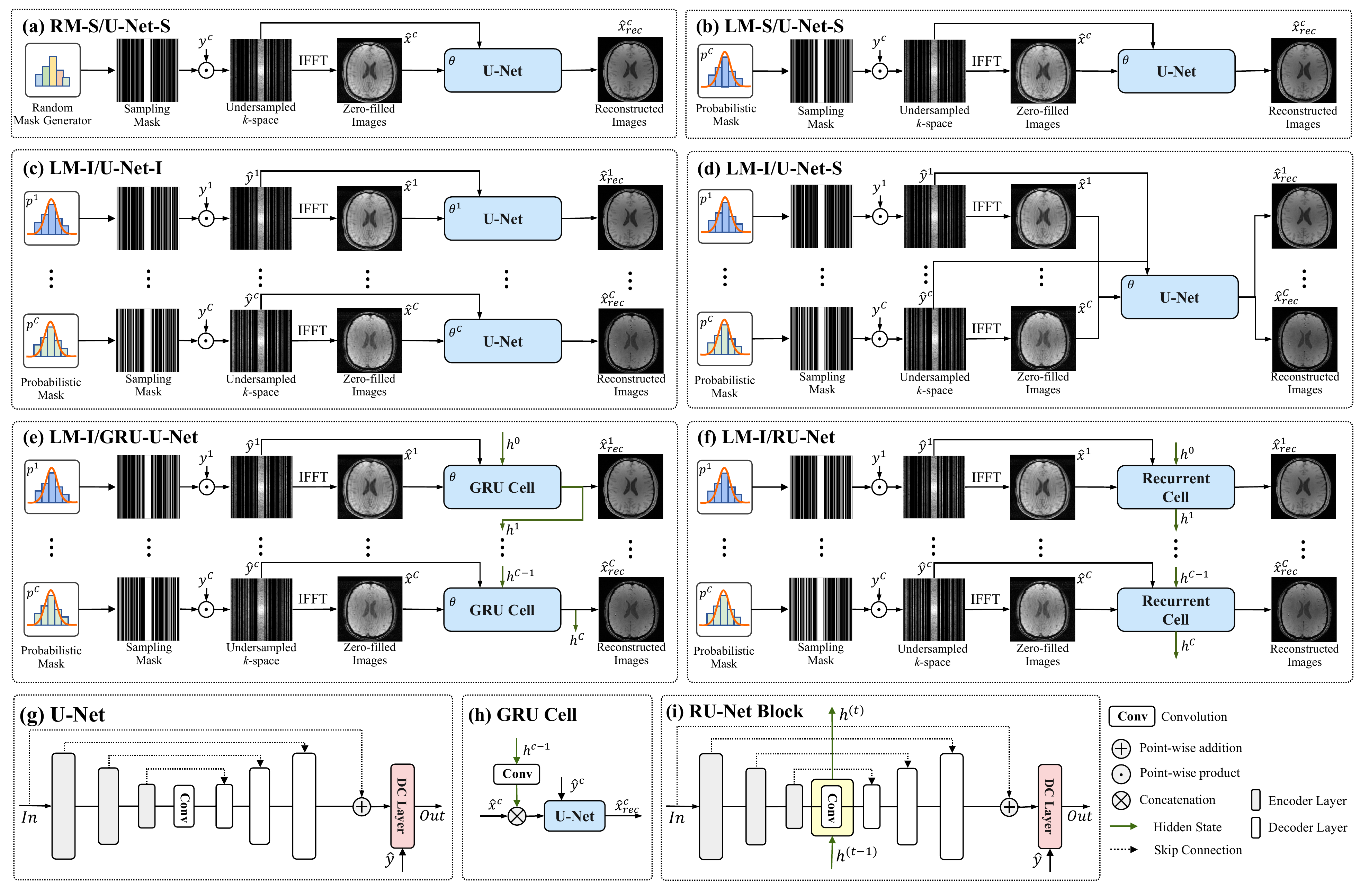}

   \caption{Illustration of baseline methods. (a)-(f) The framework architecture of six different baselines. (g) The U-Net reconstruction network used in baseline methods. (h) The architecture of the GRU Cell in LM-I/GRU-U-Net model. (i) The configuration of RU-Net block used in LM-I/RU-Net. }
   \label{fig:figureS1}
\end{figure*}

In this subsection, we elaborate the configurations of our compared baselines. The baselines are illustrated in \Cref{fig:figureS1}. 

(a) RM-S/U-Net-S consists of a random mask generator and a U-Net reconstruction network. The random mask generator produces a random sampling mask for undersampling $k$-space. The U-Net reconstruction network is composed of a standard U-Net \cite{ronneberger2015u} model and a Data Consistency \cite{schlemper2017deep} layer (\Cref{fig:figureS1}(g)), which takes the \emph{zero-filled} image $\hat{x}^c$ as input and outputs reconstruction image $\hat{x}^{c}_{rec}$. All the image contrasts use and share the same random mask generator and the same U-Net reconstruction network, i.e., different contrasts are not differentiated. 

(b) LM-S/U-Net-S adopts a learnable mask generator as in \cite{bahadir2019learning,bahadir2020deep} and a U-Net reconstruction network, which are used and shared by all image contrasts. Similar to baseline (a) RM-S/U-Net-S, the contrasts are not differentiated. 

(c) LM-I/U-Net-I, each image contrast has its own learnable mask generator and U-Net reconstruction network, which is not shared among contrasts. 

(d) LM-I/U-Net-S, each image contrast has its own learnable mask generator but the U-Net reconstruction network is shared among all contrasts. 

(e) LM-I/GRU-U-Net, each contrast has its own learnable mask generator and uses GRU-U-Net reconstruction network (\Cref{fig:figureS1}(h)) based on Gated Recurrent Unit (GRU) model \cite{cho2014learning}. In GRU-U-Net, the reconstructed image $\hat{x}^{c-1}_{rec}$ from previous contrast serves as the hidden state $h^{c-1}$ for GRU Cell. The convolution of the hidden state and the \emph{zero-filled} image are concatenated and input into a U-Net reconstruction network to produce the final reconstructed image, which is used as the hidden state $h^c$ for the next contrast.

(f) LM-I/RU-Net, each contrast has its own learnable mask generator and uses reconstruction network similar to the proposed HRU-Net model. The difference is that the HRU-Net block is replaces by the RU-Net block (\Cref{fig:figureS1}(i)) which passes the information around the bottlenetck of the U-Net as similar to \cite{wang2019recurrent}.

\section{Detailed comparisons of reconstructions}
\label{sec:compare}
In this subsection, we provide a detailed comparison of the reconstruction quality in terms of PSNR and SSIM. \Cref{tab:table1} and \Cref{tab:table2} reports the PSNR and SSIM values of all methods for each contrast and the mean values over all contrasts in multi-contrast brain dataset and fastMRI knee dataset \cite{zbontar2018fastmri}, respectively. It is observed that our proposed LM-I/HRU-Net model outperforms all compared baselines for all contrasts in both dataset, demonstrating the superiority and effectiveness of our framework. 

\begin{table*}[h]
  \centering
  \begin{tabular}{lcccccc}
    \toprule
     \multicolumn{1}{c}{Methods} & Contrast 1 & Contrast 2 & Contrast 3 & Contrast 4 & Contrast 5 & Mean \\
    \midrule
    (a) RM-S/U-Net-S & 30.78 / 0.8364 & 30.84 / 0.8295 & 31.04 / 0.8241 & 31.05 / 0.8159 & 33.09 / 0.8072 & 30.94 / 0.8226\\
    (b) LM-S/U-Net-S & 31.58 / 0.8583 & 31.41 / 0.8506 & 31.61 / 0.8442 & 31.59 / 0.8362 & 31.51 / 0.8275 & 31.54 / 0.8434\\
    (c) LM-I/U-Net-I & 31.51 / 0.8568 & 31.33 / 0.8485 & 31.60 / 0.8428 & 31.41 / 0.8332 & 31.36 / 0.8254 & 31.44 / 0.8412\\
    (d) LM-I/U-Net-S & 31.63 / 0.8576 & 31.36 / 0.8501 & 31.68 / 0.8440 & 31.53 / 0.8344 & 31.43 / 0.8258 & 31.53 / 0.8424\\
    (e) LM-I/GRU-U-Net & 31.85 / 0.8596 & 31.72 / 0.8555 & 31.89 / 0.8492 & 31.95 / 0.8425 & 31.81 / 0.8334 & 31.84 / 0.8481\\
    (f) LM-I/RU-Net & 33.12 / 0.8640 & 32.79 / 0.8583 & 33.00 / 0.8520 & 32.79 / 0.8428 & 32.58 / 0.8345 & 32.85 / 0.8503\\
    (g) \textbf{LM-I/HRU-Net} & \textbf{33.68} / \textbf{0.8711} & \textbf{33.90} / \textbf{0.8705} & \textbf{34.17} / \textbf{0.8680} & \textbf{34.04} / \textbf{0.8610} & \textbf{33.83} / \textbf{0.8553} & \textbf{33.93} / \textbf{0.8652}\\
    \bottomrule
  \end{tabular}
  \caption{The PSNR (dB) and SSIM (numbers reported in PSNR/SSIM) for all compared baselines and the proposed LM-I/HRU-Net model for each image contrast as well as the averaged values over all contrasts for multi-contrast brain dataset. Abbreviation: RM: Random Mask; LM: Learnable Mask; -S: Shared; -I: Individual.}
  \label{tab:table1}
\end{table*}

\begin{table*}[!h]
  \centering
  \begin{tabular}{lccc}
    \toprule
     \multicolumn{1}{c}{Methods} & PD & PDFS & Mean \\
    \midrule
    (a) RM-S/U-Net-S & 35.00 / 0.8990 & 32.77 / 0.8149 & 33.88 / 0.8570 \\
    (b) LM-S/U-Net-S & 36.50 / 0.9171 & 33.79 / 0.8373 & 35.14 / 0.8772 \\
    (c) LM-I/U-Net-I & 36.73 / 0.9190 & 33.66 / 0.8359 & 35.20 / 0.8775 \\
    (d) LM-I/U-Net-S & 36.73 / 0.9190 & 33.70 / 0.8365 & 35.22 / 0.8777 \\
    (e) LM-I/GRU-U-Net & 36.85 / 0.9202 & 33.74 / 0.8377 & 35.29 / 0.8789 \\
    (f) LM-I/RU-Net & 38.12 / 0.9229 & 34.00 / 0.8406 & 36.06 / 0.8853 \\
    (g) \textbf{LM-I/HRU-Net}  & \textbf{38.48 / 0.9329} & \textbf{34.15 / 0.8441} & \textbf{36.32 / 0.8885}  \\
    \bottomrule
  \end{tabular}
  \caption{The PSNR (dB) and SSIM (numbers reported in PSNR/SSIM) for all compared baselines and the proposed LM-I/HRU-Net model for each image contrast as well as the averaged values over all contrasts for the multi-contrast fastMRI knee dataset. Abbreviation: RM: Random Mask; LM: Learnable Mask; -S: Shared; -I: Individual.}
  \label{tab:table2}
\end{table*}

\section{Downstream task T2* map synthesis}
\label{sec:learn}
In this subsection, we provide detailed descriptions and results of the downstream task. T2* map synthesis is used as an example for downstream task. On the T2*-weighted multi-contrast images, the pixel value on the $c^{th}$ image at location $(u,v)$ is an exponential decay signal with time constant $T2^{*}(u,v)$: $s(u,v,c)=e^{-c\cdot\Delta t/T2^{*}(u,v)}$, where $\Delta t$ is a constant unit time. Assuming $f(\cdot)$ is a function that maps $\{s(u,v,1), s(u,v,2), ..., s(u,v,C)\}$ to $T2^{*}(u,v)$, a complete T2* map is synthesized after $f(\cdot)$ is applied at all pixel locations. We used a neural network as a regressor with three fully connected layers that have 5/32/1 nodes, respectively. To train the network, 64 million T2*-decay signals with uniformly random T2* ranging from 0 to 300 ms which covers the practical T2* values in human brain MR imaging are simulated. Mean square error loss is used for training the regressor and the training is converged after 1 million steps. After training, the network parameters are fixed and not updated for other experiments. The neural network based $f(\cdot)$ allows backpropagation of the loss of the T2* map to update sampling and reconstruction parameters. Please note that the main goal of the design of T2* map synthesis is to test the capability of the proposed method to optimize data sampling and reconstruction across different contrasts when given a task where different contrasts contribute unequally. The accuracy of the regressor trained for T2* map synthesis is not the focus here.

\begin{figure*}[t]
  \centering
   \includegraphics[width=1\linewidth]{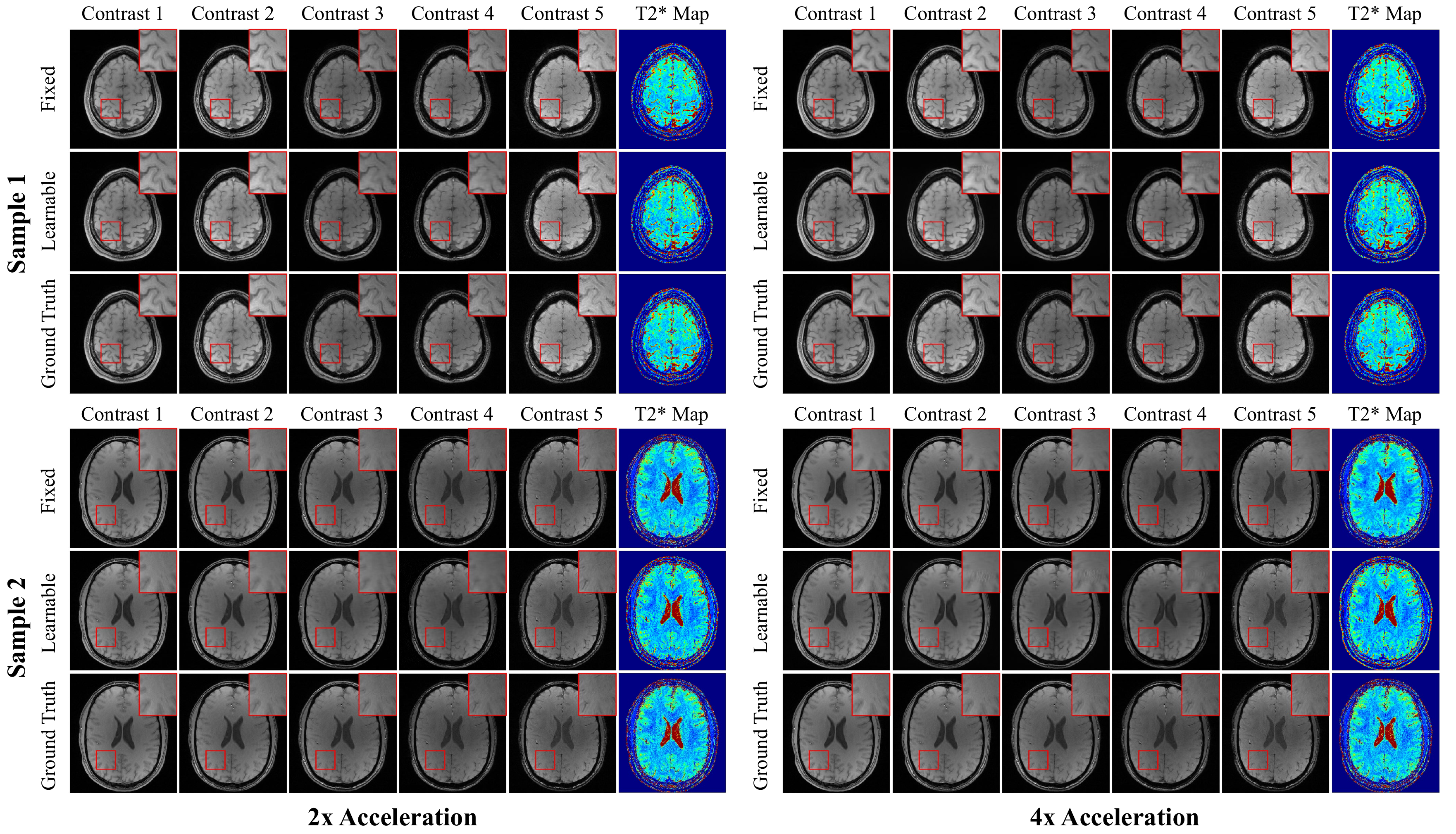}

   \caption{The reconstructed images and the synthesized T2* maps based on fixed and learnable acceleration ratio under 2x and 4x acceleration.}
   \label{fig:figureS3}
\end{figure*}

\Cref{fig:figureS3} shows multiple examples of the reconstructed images of each contrast and the final synthesized T2* map from fixed/learnable acceleration and ground truth. As reported in the main manuscript, the learnable acceleration ratio training result has high acceleration ratios for contrast 2-4 while maintains low acceleration ratios for contrast 1 and 5. Correspondingly, the image quality is poorer for contrast 2-4 and better for contrast 1 and 5 when compared to those from fixed acceleration ratio. It is noted here that the sampling and reconstruction is optimized only on the downstream task, which is the T2* map quality. As shown here, with a fixed mapping function, better image quality for some of the contrasts does not translate to better map quality. What's more, for an exponential decay fitting task such as the T2* map synthesis, two data points are sufficient. It is beneficial for the fitting accuracy and robustness that the first and the last contrasts are of high quality, since that pair gives the biggest dynamic range. The learned acceleration ratios are reasonable and agree with this prior knowledge. 
We report PSNR and SSIM of the T2* map with background and without background. To remove the background, a brain foreground mask is calculated from the multi-contrast brain MR images. Specifically, the magnitude of the multi-contrast images are averaged and then a Sobel mask is calculated followed by thresholding to binarize the image. The binary mask is dilated and holes are filled. Finally, the largest connected component is kept which covers the whole brain. Please note that the mask is only used for evaluation.

\newpage
\newpage
%%%%%%%%% REFERENCES
{\small
\bibliographystyle{IEEEtran}
\bibliography{egbib}
}